\documentclass[superscriptaddress,amssymb,showpacs,a4,twocolumn]{revtex4}

\usepackage{epsfig}
\usepackage{graphics}
\usepackage{graphicx}

\begin{document}

\title{Non-linear mechanical response of the Red Blood Cell}

\author{Young-Zoon Yoon}
\affiliation{Cavendish Laboratory and Nanoscience Center, University
of Cambridge, Cambridge CB3 0HE, U.K.} \affiliation{Institute for
Biomedical Electronics, Seoul National University of Technology,
Seoul 139-743, Korea}

\author{Jurij Kotar}
 \affiliation{Cavendish Laboratory and
Nanoscience Center, University of Cambridge, Cambridge CB3 0HE,
U.K.}

\author{Gilwon Yoon}
 \affiliation{Institute for Biomedical Electronics, Seoul National
University of Technology, Seoul 139-743, Korea}

\author{Pietro Cicuta}
 \affiliation{Cavendish Laboratory and
Nanoscience Center, University of Cambridge, Cambridge CB3 0HE,
U.K.}
 %

\begin{abstract}
We measure the dynamical mechanical properties of human red blood
cells. Single cell response is measured with optical tweezers. We
investigate both the stress relaxation following a fast deformation,
and the effect of varying the strain rate. We find  a power  law
decay of the stress as a function of time, down to a plateau stress,
and  a power law increase of the cell's elasticity as a function of
the strain rate. Interestingly, the exponents of these quantities
violate the linear superposition principle, indicating a nonlinear
response. We propose that this is due to breaking of a fraction of
the crosslinks during the deformation process. The Soft Glassy
Rheology Model accounts for the relation between the exponents we
observe experimentally. This picture is consistent with recent
models of bond remodeling in the red blood cell's molecular
structure. Our results imply that the blood cell's mechanical
behavior depends critically on the deformation process.
\end{abstract}


\maketitle


The human red blood cell is a biological structure of relative
simplicity: it lacks a nucleus and intra membrane organelles. Its
components are well characterized~\cite{alberts94,boal02,lim02}, and
they can be summarized as a bilayer membrane, coupled to a thin
tenuous cytoskeleton of spectrin filaments via two complexes of a
few different proteins (ankyrin, band 3 and band 4.1).  This outer
membrane, held under tension by the cortical cytoskeleton,  encloses
a solution of dense haemoglobin. Despite its simplicity, the
properties of this structure have puzzled researchers for decades.
Investigations have focussed on two, related aspects: explaining the
biconcave discocyte shape that is found under physiological
conditions (and a whole array of other morphologies that are found
under perturbation or disease); measuring the mechanical properties
of the structure. Regarding the shape, significant progress has been
made recently showing how, by carefully balancing the membrane
bending elasticity with a tension from the underlying spectrin
scaffold, the discocyte shape emerges as the equilibrium solution,
and the cup-shaped stomatocyte is the result of a perturbation that
decreases the membrane area relative to the underlying
skeleton\citep{lim02}.  On the issue of mechanics and dynamics of
the structure, our understanding is still very limited, despite the
great importance of these parameters. Large deformations of the cell
are involved   in blood flow through thin capillaries,  and the
cells in suspension more generally determine the peripheral blood
rheology in healthy and diseased conditions~\cite{suresh06}. The red
blood cell has variously been described as either
liquid or solid~\cite{hochmuth82}.\\
In this Letter we investigate the dynamical mechanical properties of
the red blood cell by measuring its viscoelasticity over a range of
environmental conditions. We focus on  the effect of the cell age,
of the deformation dynamics, and the correlation between the cell
modulus and its shape. We relate the findings to phenomenological
models for flow in kinetically arrested systems, and to recent
models describing the metabolic activity of the RBC, where the
consumption of ATP controls the stiffness of the elastic network
\citep{korenstein92,korenstein98,gov07b}. The normal biconcave human
RBC has approximately 25\% more membrane surface area than the
minimum required to enclose its volume, allowing deformations
without extension of the membrane~\cite{jay75}. Therefore the
mechanical response of the cell is related only to its bending and
shear elastic moduli, and its membrane and bulk viscosities. These
quantities have been measured with various techniques, in particular
micropipette aspiration, shape flickering, deformation in flow,  and
optical tweezers. Existing measurements give very different values.
For example,  independent experiments using micropipette aspiration
yield a shear modulus of between 6 and 10$\mu$N/m~\citep{evans76}.
Optical tweezers measurements report a wide range of shear modulus,
but the discrepancies between these
 results are in part the result of applying
different geometrical models to extract the  modulus from the
measured forces: 2.5$\pm$0.4$\mu$N/m assuming a flat disk
geometry\cite{henon99}; $200\mu$N/m assuming a spherical initial
shape \citep{sleep99,winlove99}; 11 to 18$\mu$N/m comparing to a
finite element simulation of the deformation~\cite{suresh03}. A
satisfactory approximation for the deformation geometry is still not
available, and is not the topic of this Letter. We choose instead to
report the directly measured mechanical properties, showing that
even the model-independent dynamical stiffness of the cell is
strongly dependent on the deformation protocol. Our key result is
that the RBCs response cannot be explained within the framework of
linear viscoelasticity.\\
\begin{figure}[b]
\centering
\begin{tabular}{cc}
           \epsfig{file=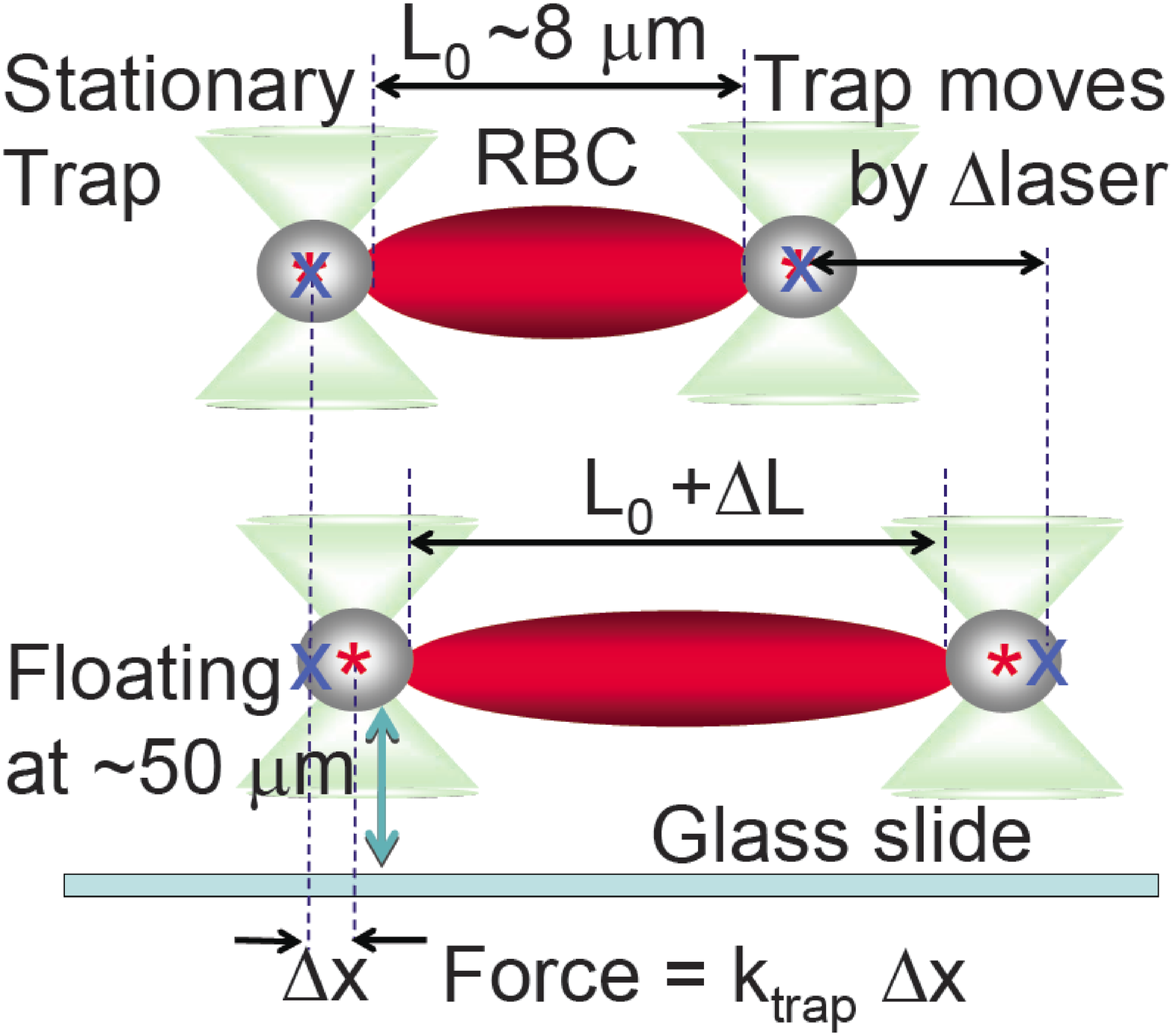,width=3.5cm} & \epsfig{file=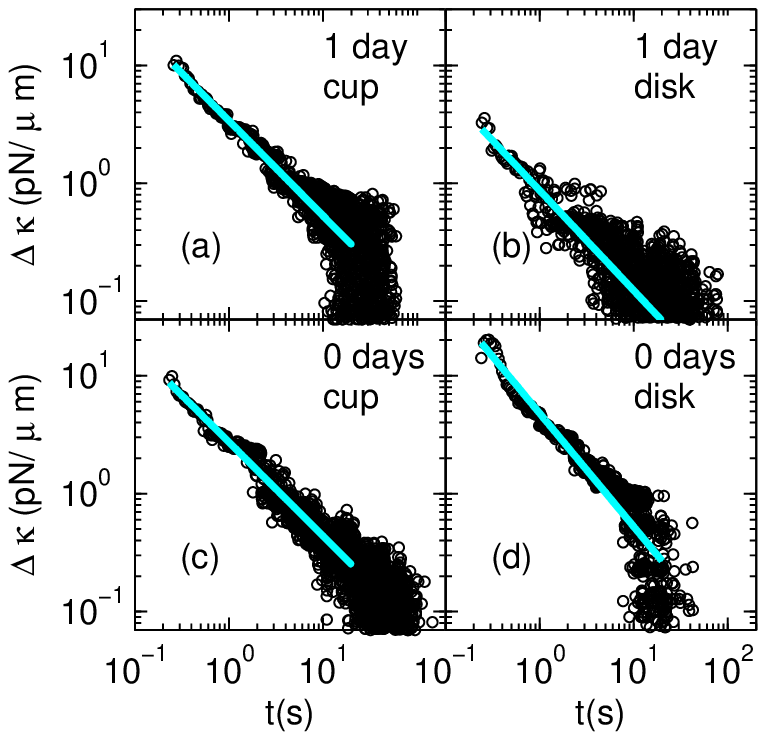,width=5cm}
            \end{tabular}
 \caption{Diagram of experimental setup. A cross marks the positions of the laser trap and
  a star marks the center of the bead, obtained via image analysis.
 (a)-(d)~Decays of the measured time-dependent cell stiffness $\kappa (t)$ as a function of time.
 The plateau value has been subtracted. This data
 shows a power law decay with exponents $\simeq$3/4.
  \label{fig1}}
\end{figure}
Fresh blood was drawn from a healthy volunteer donor, and diluted in
phosphate-buffered saline (PBS) with acid citrate dextrose (Sigma
C3821) and 1mg/ml bovine serum albumin (BSA) (Sigma A4503) at pH7.4.
For studying the effect of aging time after drawing, cells were
incubated with glucose-free PBS either in a hot bath plate at
37$^o$C for 24h with 10$^{-2}$ Penicillin-Streptomycin solution
(GIBCO, Invitrogen 15070063) to prevent  microbial growth, or at
4$^o$C for 24h.   A first control group was formed by adding
1mM~MgATP (Sigma A9189), after the 24h incubation, to the the RBCs
in glucose-free PBS suspension. A second control group was incubated
with 100mg/dl glucose in PBS buffer. These protocols provide samples
that are a mixture of discocyte and stomatocyte cells, with an
increasing fraction of
discocytes after 1 day's incubation.\\
  The optical tweezers setup consists of a laser (IPG Photonics,
PYL-1-1064-LP, $\lambda$=1064nm, P$_{max}$=1.1W) focused through a
water immersion objective (Zeiss, Achroplan IR 63x/0.90 W) trapping
from below. The laser beam is steered via a pair of acousto-optic
deflectors (AA Opto-Electronic, AA.DTS.XY-250@1064nm) controlled by
custom built electronics allowing multiple trap generation with sub
nanometer position resolution.  The trapping potential is locally
described by a harmonic spring, and  the trap stiffness was
calibrated
 by measuring the thermal displacements of a
trapped bead.  The trapping stiffness at maximum power when trapping
two beads (as in this work) is $k_{\textrm{trap}}$=44pN/$\mu$m on
each bead. The sample is illuminated with a halogen lamp and is
observed in bright field with
a fast CMOS camera (Allied Vision Technologies, Marlin F-131B).\\
   Carboxylated silica beads of 5.0$\mu$m diameter (Bangs
Labs) were washed  in Mes buffer (Sigma M1317). They were then
functionalized by resuspending in Mes buffer and incubating for 8h
at 37$^o$C with Lectin (0.25mg/ml) (Sigma L9640) and EDC (Sigma
E1769) (4mg/ml). This made them very sticky to the RBC, probably
through binding to glicoproteins or sugars on the outer side of the
RBC membrane. To prevent the RBCs from sticking on the glass
surfaces of the chamber,  BSA  was coated on the slide glass. Using
the tweezers, two beads are brought to an RBC and attached
diametrically across a cell, which is then floated well above the
glass slide surface (at about 10 times the bead diameter) to
minimize any hydrodynamic drag from the solid surface. The cells are
quite monodisperse in size, with an initial cell length (the
diameter) L$_0\sim8\mu$m. The area of contact between the bead and
the cell varies in the range 3-4.5$\mu$m$^2$, and we find no
correlation of the patch area with any of the experimental results.
This arrangement is drawn in Figure~\ref{fig1}. During all the
deformation protocols, one laser trap is  kept fixed at its initial
position, while the other trap is moved away by a distance
$\Delta$laser. We focus on tracking the bead in the immobile trap,
measuring the difference $\Delta$x between the bead and the known
laser position using image analysis code written in Matlab. From the
displacement $\Delta$x, the stretching force is calculated. The cell
is elongated by a distance $\Delta \textrm{L}=\Delta \textrm{laser}-
2\Delta \textrm{x}$, and we define the strain as $\gamma=\Delta
L/L_0$. The resolution of bead position via image analysis on each
frame is around 5nm, which translates into $\pm$0.22pN of force
resolution. This is significantly less than the random fluctuations
caused by thermal noise. The cell stretching is recorded at $\sim$60
frames per second, and having checked that we could not observe a
variation between measurements done at room temperature and at
37$^o$C, we report on measurements made at room temperature.
Temperature is known
from previous studies  to have only a slight influence on the mechanics of red blood cells~\cite{sackmann84,artmann95}.\\
We perform stress relaxation experiments by moving just one trap
(i.e. moving one bead) by $\Delta$laser=3.5$\mu$m at 20$\mu$m/s, and
monitoring the force acting on the stationary bead. The force is
observed to decay towards a plateau value, which is reached within
less than half a minute. We define the time-dependent stiffness of
the red blood cell to be $\kappa (t)= F(t)/\Delta \textrm{L}(t)$.
The whole set of experiments is fit very well by the 3-parameter
power law function:
\begin{eqnarray}
  \kappa(t)\,=\,\kappa_\infty \,+\,\Delta \kappa(t)\,=\,\kappa_\infty \,+\,\Delta \kappa_0\,(t/t_0)^{-\alpha},
 \label{eq1}
\end{eqnarray}
where we fix $t_0=1$s to obtain a dimensionless time. We use the
first 20 seconds of the decay to fit these values, where there is
less noise in the data, but it can be seen that the power law decay
holds for longer times as well. In contrast, a single exponential
fails to fit the data and even a stretched exponential
(4-parameters) gives a poor fit. Figure~\ref{fig1} shows examples of
the time relaxation of $\Delta \kappa(t)$ plotted separately for
each of discocyte and cup shaped stomatocyte cells, fresh and after
one day of incubation. In all cases the stiffness is fit over two
decades in time by the form of Eq.~\ref{eq1}. The power law
exponents are the same within experimental error, the mean of all
the power law exponents measured is $\alpha$=0.75$\pm$0.16 for fresh
cells and $\alpha$=0.82$\pm$0.09 for 1-day old cells. A total of 11
different cells were measured this way. The value of the ratio
$\Delta F_0/F_\infty$, i.e. the fraction of the stress that decays
over time, also appears independent of shape and cell age, and is
between 0.55 and 0.75.  The fact that the data is  described so well
by Eq.~\ref{eq1} implies that there are two components to the cell
elastic modulus: a time-dependent and a long-time equilibrium
modulus. Power-law relaxations are well known in systems near the
gelation point and in entangled polymer solutions~\cite{larson99},
however we anticipate here that we will discuss how for the red
blood cell the form of Eq.~\ref{eq1} does not originate from polymer
relaxation nor gel cluster dynamics.
\begin{figure}[b]
           \epsfig{file=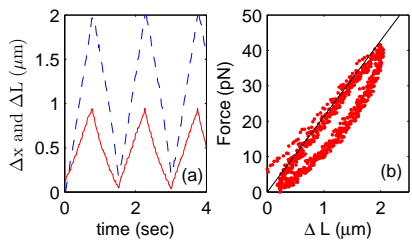,width=5.5cm}
 \caption{(a) The dashed line is the elongation $\Delta$L of the red blood
 cell subject to a  triangle wave trap displacement, and the solid
 line is $\Delta$x of the bead in the stationary trap.  (b)~Force vs. elongation of the cell $\Delta$L, for
 the same data  as in panel (a). Solid line is a linear fit to the force during extension,
  the area inside the  hysteresis loop  is the energy dissipated per cycle.
  \label{fig3}}
\end{figure}
To investigate further the strain rate dependence of the cell
modulus, we perform a triangle wave deformation experiment using
speeds of  20, 5, 1 and 0.2$\mu$m/s,  and 3.5$\mu$m laser trap
displacement \footnote{For one speed and condition  (data not shown)
we applied a triangle wave of amplitudes 1, 5 and 7 $\mu$m and
obtained the same stiffness as for 3.5$\mu$m.}. Figure~\ref{fig3}(a)
shows that the force measured over time is
 approximately proportional to the applied strain. The
slight mismatch between the applied strain and the stress can be
seen more clearly in Figure~\ref{fig3}(b) as a hysteresis loop.
There is  a slight difference  between the onset of the first cycle
and all the following ones: From observing many instances of this
protocol we believe that this is  due to the cell rearranging its
position precisely along the axis of the strain experiment. It is
clear from this figure that there is no further evolution of the
stress strain curves over the run, which lasts typically tens of
cycles. The cell's response is predominantly elastic, and the
gradient of the force vs. elongation data as in
Figure~\ref{fig3}(b), taken over the extension, gives  the stiffness
$\kappa$ of the cell. This
 stiffness is an effective property of the cell, a valid modulus for describing the cell's  response
 in physiological processes such as squeezing
through thin capillaries. Our results show that
  there can be a large (factor of three) difference between the dynamic and static
  stiffness, as measured for example by micropipette.  The experimental speed of deformation can
be converted into an approximate strain rate $\dot{\gamma}$ by
dividing the bead velocity by the initial cell length (L$_0$). The
cell stiffness is plotted in Figure~\ref{fig4} against the strain
rate. These quantities are seen to be approximately related through
the function:
\begin{eqnarray}
  \kappa(\dot{\gamma})-\kappa_0\,\sim\,\dot{\gamma}^\beta,
 \label{eq2}
\end{eqnarray}
with $\beta\simeq\frac{1}{4}$.\\ The hysteresis in the stress strain
curves is a measure of energy dissipation in the cell, and is shown
in Figure~\ref{fig4}.    In order to compare results from different
strain amplitudes, the dissipated energy is presented as the work
per cycle, normalized by the square of the  strain amplitude
$\gamma_{max}$. The dissipated energy  follows a power
 law as a function of strain rate, with an exponent that we assume is the same $\beta$ as in Eq.~\ref{eq2}.
    In all the cells
the strain rate dependence is well below the linear relation
expected for a purely viscous system.   The exponents fitted here
are $\beta=$0.25 for fresh discocytes and
 $\beta=$0.33 for  discocytes after one day's incubation. These values have a higher precision than
 the stress relaxation exponents because the triangle wave cycle is repeated many times.      \footnote{We
 consider here how much dissipation can arise
 from the phospholipid bilayer itself. Its viscosity is around 0.3Pa\,s,
 ref.~\cite{cicuta07a}. For a simplified geometry of
deformation of two flat plates, the dissipated energy is around
0.6$\times10^{-18}$J per cycle at $\dot{\gamma}=10^{-2}$s$^{-1}$.
This is an order of magnitude below our results but, growing
linearly with strain rate, it would dominate the response at higher
deformation rates.
  }
  In linear viscoelasticity, the
dependence of the modulus on the strain rate is related to the form
of the stress relaxation  via a modified Laplace
transform~\cite{larson99}. In the case of a power law decay with
$\kappa(t)-\kappa_\infty \sim t^{-\alpha}$, we expect
$\kappa(\dot{\gamma})-\kappa_0\sim \dot{\gamma}^\alpha$. This is
very clearly not the case here, the exponent $\beta$ is much smaller
than $\alpha$. This points to the fact that even though our strain
amplitudes are not very large, only of the order of 20\%, our
experiment is in a nonlinear regime. This finding can be understood
as follows: the power law  relaxation is the manifestation of a
system where at least a fraction of the bonds can break under stress
and quickly reform, essentially remodeling the skeleton's network.
There is an analogy between this molecular-scale picture and the
more general situation of the rheology of a system where the
components have to overcome potential barriers in order to flow.
This condition is addressed by the soft glassy rheology (SGR) model
which, assuming a distribution of energy wells of different depths,
provides a framework for calculating the nonlinear response of the
system\cite{cates97}. It contains one principal parameter, $x$,
which is the ratio between the available energy and the mean well
depth. For $x<1$ the model is in a glassy phase. The shear modulus
is predicted to decay with time as $G \sim t^{-x}$, and the stress
under constant strain rate is $\sigma-\sigma_y \sim
\dot{\gamma}^{1-x}$. The red blood cell power laws and exponents
appears to be well described by the SGR model with a value of
$x\simeq 3/4$. In the red blood cell it is likely that the potential
barriers to be overcome are the energies for releasing a spectrin
filament from a crosslink. The filament would then re-bond in a
configuration of lower stress. The presence of a long lived residual
stress in the stress relaxation
 implies either the presence of a fraction of permanent
bonds, or of a kinetically arrested state with residual stress,
leading to a deviation from a purely power law scaling at low strain
rates which is visible in the stiffness
data of Figure~\ref{fig4}.\\
\begin{figure}[t]
           \epsfig{file=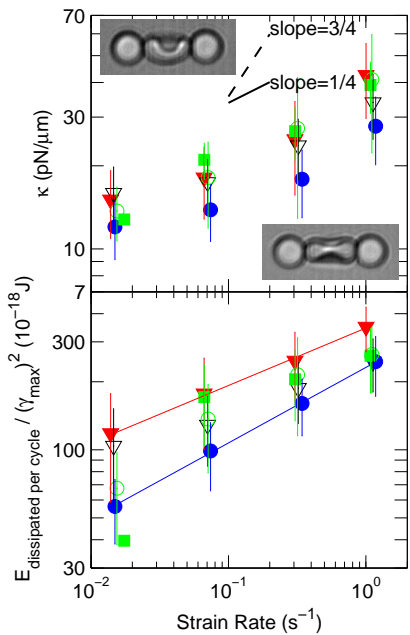,width=5.5cm}\\
 \caption{Strain rate dependence of the RBC stiffness $\kappa$ and the energy
 dissipation. Markers
 indicate the different conditions studied here:
 ($\blacktriangledown$)~fresh-Discotic;
 ($\triangledown$)~fresh-Stomatocyte;
 ($\bullet$)~1day-Discotic; ($\circ$)~1day-Stomatocyte;
 ($\blacksquare$)~1day-discotic +MgATP.
  \label{fig4}}
\end{figure}
 In the absence of external stresses it is known that ATP is required to open the
spectrin-actin linker, whose bond energy
 is  around 7k$_B$T, i.e. 3$\times
10^{-20}$J~\cite{bennett89,gov05}. Based on the known mesh size of
the cytoskeleton (between 80-100nm) and the cell surface area, we
can estimate the order of magnitude of the number of bonds to be
10$^4$. Given that the dissipated energy per cycle is of the order
of $10^{-17}$J this  implies  that either only a small fraction of
the bonds, around 3\%, is broken during a cycle, or that  the bond
energies under stress are reduced considerably, or that ATP and not
solely mechanical energy is being used to break the bonds. Of these
possibilities the agreement with the soft glassy rheology observed
here suggests that remodeling of the linkers under stress
does not involve the full cost of breaking the spectrin-actin linker. \\
While the features of the dynamical response of all the cells are
the same, we observe differences in the magnitude of the response
depending on the shape  of cell and the incubation protocol. The
strongest influence,  leading to a decrease of the cell modulus of
almost a factor of 2 after 1 day of incubation, is
 the time from when the cells are drawn.  Secondary to this aging,  we
observe a correlation between the stiffness of the cell and its
shape. The cup-shaped stomatocytes are almost twice as stiff as the
discotic cells.    Models proposed recently describe the active
remodelling of the RBC's cytoskeletal network~\cite{gov05,gov07b}.
ATP is expected to be strongly related to this activity, at least
under absence of an external force~\cite{gov07b}. It is of interest
that flickering of the membrane at low frequencies was shown to
depend on the intracellular MgATP\cite{korenstein91}, and the
flickering has also been shown to decrease with cell
aging~\cite{sackmann84}. Through the 1-day incubation protocol
 the ATP concentration inside the cell is depleted, and according to these references
 this would lead to stiffer
 cells. Our results showing softening
 seem to imply an opposite effect, one possibility is that under reduction of ATP the cell
  loses some of its ability to
 reform the bonds that break under deformation. We find  no appreciable
difference between aging in the absence of glucose and in
physiological glucose concentrations (data not shown). We find a
strong effect adding MgATP to an aged cell: MgATP causes an increase
of both the stiffness and dissipation for the discotic cells,
bringing the values close to
those of freshly drawn cells. \\
The rate dependence of the cell modulus measured in this work can be
also be compared to recent data obtained by monitoring the
deformation induced via magnetic twisting cytometry (MTC). This is
essentially a creep experiment performed on a pivoting magnetic bead
bound to the outside of a red blood cell~\cite{suresh07}. In MTC
there is almost certainly a large influence of the  membrane bending
elasticity, and the data in ref.~\cite{suresh07} is dominated by a
source of elasticity that is not found in our measurements with
optical tweezers. Our data confirms that the dynamics in the red
blood cell, as also pointed out by Suresh and collaborators, is
intrinsically free of a characteristic timescale~\cite{suresh07}. In
addition we have shown here that the dynamics is not linear, and
therefore  depends on the deformation protocol, even for small
deformations. This may be relevant to  other cell types. Our
findings here are in contrast to many previous investigations of the
red blood cell~\cite{evans79,henon99,suresh03}, where limited
datasets led to simplistic conclusions regarding the dynamics.\\
%
%
%
%
%
%
%
%
%
%
%
We acknowledge funding by the Oppenheimer Fund, EPSRC, and the
Cavendish-KAIST programme of MoST Korea.  We thank  I. Poberaj, N.
Gov, E.M. Terentjev, J. Guck, J. Sleep, J. Evans, W.Gratzer and P.G.
Petrov for advice and help.

\bibliography{bibdatav21}

\end{document}